# Enhancement of magnetodielectric coupling in 6H-perovskites Ba₃$R$Ru₂O₉ for heavier rare earth cations ($R$=Ho,Tb)

*Tathamay Basu,[1,2,3]\* Vincent Caignaert, [1] Somnath Ghara [3] Xianglin Ke,[2] Alain Pautrat,[1] Stephan Krohns,[3] Alois Loidl[3] and Bernard Raveau,[1]*

*[1]Laboratoire CRISMAT, UMR 6508 du Cnrs et de l'Ensicaen, 6 Bd Marechal Juin, 14050 Caen, France*

*[2]Department of Physics and Astronomy, Michigan State University, East Lansing, Michigan 48824, USA*

*[3]Experimental Physics V, Center for Electronic Correlations and Magnetism, University of Augsburg, Universitätsstr. 2, D-86135 Augsburg, Germany*

[\*tathamaybasu@gmail.com](mailto:tathamaybasu@gmail.com)

Keywords: Magnetism, Ferroelectric, Multiferroic, Magnetodielectric, 4$d$-4$f$ interaction

## Abstract

**The role of rare-earth ($R$) ions on the magnetodielectric (MD) coupling is always intriguing and markedly different for different systems. Although many reports are available concerning this aspect in frustrated 3$d$-transition metal oxides, no such reports exist on higher $d$ (4$d$/5$d$)-orbital based systems due to the rare availability of highly insulating 4$d$/5$d$-systems. Here, we systematically investigated the magnetic, dielectric, ferroelectric and magnetodielectric behavior of the 6H-perovskites Ba₃$R$Ru₂O₉ for different $R$-ions, namely, $R$= Sm, Tb and Ho, which magnetically order at 12, 9.5 and 10.2 K respectively. For $R$=Tb and Ho, the temperature and magnetic-field dependent complex dielectric constant traces the magnetic features, which manifests MD coupling in this system. A weak magnetic-field ($H$) induced transition is observed for ~30 kOe, which is clearly captured in $H$-dependent dielectric measurements. No MD coupling is observed for Ba₃SmRu₂O₉. The MD coupling is enhanced by a factor of 3 and 20 times for R=Tb and Ho, respectively, when compared to that of the Nd-counterpart. These results evidence the gradual enhancement of MD coupling with the introduction of heavier $R$-ions in this series, which is attributed to their larger moment values. Our investigation establishes dominating 4$d$(Ru)-4$f$($R$) magnetic correlation in this series for the heavier $R$-members.**

## I.    Introduction:

The coexistence of ordered magnetic and lattice degrees of freedom and cross-coupling between them have been attracting intense attention due to the fundamental scientific interest and promising applications in future storage devices.[1–3] After the discovery of magnetism-induced ferroelectricity in TbMnO₃,[4] it has been found that many frustrated magnetic systems, containing 3$d$-metal-ions and R-ions, exhibit multiferroicity and spin-dipole coupling.[1,5,6] In all these systems, both $R$-ions ($f$-orbital) and transition metals ($d$-orbital) play an important role in establishing magnetism and concomitant magnetoelectric (ME)/ magnetodielectric (MD) coupling. The exact role of R-ions on ferroelectricity and ME/MD coupling has not been finally clarified and still is a matter of debate. In general, the ME coupling strength should directly depend on the strength of the polarization, magnetization and the coupling constant of that system associated with different mechanisms. The different size of *R*-ions (lanthanide contraction) may play a decisive role on lattice distortion (bond length, angles, etc.) of the system, thereby directly affecting dielectric properties. Further, different radii of *R*-ions (different degree of localization/ hybridization) and the large magnetic moments of *R*-ions should have an effect on the overall magnetic structure (e.g. change in exchange interactions and/or heavy moment of *R*-ion could affect the canting angle of magnetic moment of transition metal (TM) ion).

It has been demonstrated that the interactions between the *3d* and *4f* electrons of the TM and *R* sites, respectively, have an important role in their magnetism and magnetoelectric coupling. This is exemplified by the multiferroic *R*MnO$_3$ perovskites where the size of the *R*-ion directly affects the Mn-O-Mn angle and thereby modulates the magnetic structure.[7] For orthorhombic (distorted perovskites) *R*MnO$_3$, (*R*= Eu, Gd, Tb and Dy), an incommensurate cycloidal magnetic structure of Mn is observed, which breaks the inversion symmetry (as a result of asymmetric Dzyaloshinskii–Moriya (D-M) interaction ) and drives ferroelectric polarization.[8] A giant MD coupling is observed in DyMnO$_3$ and GdMnO$_3$ compared to other R-members.[8] The effect of symmetric exchange-striction, in addition to asymmetric D-M interaction, is reported for *R*= Gd and Dy members in this series.[9] Further heavy rare-earth members crystallize in an hexagonal structure (e.g., HoMnO$_3$) and exhibit commensurate antiferromagnetic (AFM) ordering. The ferroelectricity in the hexagonal HoMnO$_3$ oxide arises as a result of displacements of Ho-ions and a possible tilting of MnO$_5$ polyhedra, and the magnetic structure is not directly involved to create ferroelectricity.[10,11] The MD coupling in this hexagonal system originates from magnetoelastic coupling, where magneto-striction plays an important role. *R*Mn$_2$O$_5$ oxides constitute a second example of multiferroic materials, whose magnetoelectric properties are governed by *3d-4f* electron interactions.[12–16] In *R*Mn$_2$O$_5$, the presence of loops of 5 manganese MnO$_5$/MnO$_6$ polyhedra sharing corners and edges (Mn$^{3+}$-Mn$^{4+}$- Mn$^{3+}$-Mn$^{4+}$-Mn$^{3+}$) gives rise to magnetic frustration. It is considered that the symmetric exchange-striction arising from this frustrated structure creates off-centering of Mn$^{+3}$-ion, *R* (4*f*) –Mn (3*d*) coupling plays an important role on this exchange-interaction. The ME coupling is stronger for DyMn$_2$O$_5$ compared to that of TbMn$_2$O$_5$, whereas TbMn$_2$O$_5$ exhibits a switching of polarization in low magnetic fields.[12,14] Interestingly, GdMn$_2$O$_5$ exhibits high polarization and ME coupling compared to all other *R*-members in this series.[15] The investigation of the light rare-earth (*R*= Pr, Nd) members reveals weak ME coupling compared to the heavy *R*-members.[16] Another series of *R*CrTiO$_5$, derived from the similar *R*Mn$_2$O$_5$ structure, also exhibits multiferroicity and ME coupling, though the role of *R*-ion in this series is different to that in RMn$_2$O$_5$.[17–19] The ME coupling of NdCrTiO$_5$ is stronger than that of GdCrTiO$_5$,[17,19] unlike the ME effect characteristics for the RMn$_2$O$_5$ series. The Haldane-chain oxides, *R*$_2$BaNiO$_5$, form another family, where strong *3d-4f* correlation exist via Ni-O-*R*-O-Ni super-exchange paths, and recently have received considerable attention with respect to multiferroicity and ME coupling.[20–25] In this series, magnetism-induced ferroelectricity with strong ME coupling is proposed for Dy$_2$BaNiO$_5$,[21] whereas ferroelectricity is observed at temperatures well above long-range ordering for many other members (*R*= Ho, Er, etc.).[20,22] The light rare-earth Nd-member does neither exhibit ferroelectricity nor ME coupling.[26] A displacement- type ferroelectricity (from NiO$_6$ distortion in Ni spin-chain) is proposed for the compound $Er_2BaNiO_5$.[22] Interestingly, the compound $Tb_2BaNiO_5$ exhibits a giant MD coupling,[24] where it is reported that D-M interaction between two different magnetic ions (Tb and Ni) is responsible for ferroelectricity, suggesting the influence of strong $R$ (*4f*) –Ni (*3d*) magnetic correlations on ferroelectricity and ME coupling.[25] Therefore, it is quite clear that *3d-4f* interaction is complex depending on the *R*-ion (different size of *4f*-orbital and magnetic moment) and it behaves completely different in various systems.

Till now, the research on multiferroicity/ ME coupling in oxides has been mainly restricted to *3d*-TM cations. In the past decade, there has been tremendous interest in *4d* or *5d* TM oxides due to their exotic magnetic behavior, arising from the extended electronic orbitals, crystal-field effects, and strong spin-orbit coupling, such as, Mott or topological insulating behavior, a quantum spin-liquid state, unconventional superconductivity or field-driven insulator-metal transitions (for instance, see [27–34] and references there in). However, there are only very few studies of MD coupling in compounds containing higher *d*-orbital of (*4d/5d*) TM-ions and *4f*-orbital *R* ions, despite the high interest from theory.[35] Though the larger extension of the *4d* or *5d*-TM orbitals should favor stronger interactions with the *4f-R* orbitals, it also induces larger overlapping of *4d-4d* (or *5d-5d*) orbitals via oxygen and consequently the investigation of such systems is often experimentally hindered due to the leaky nature (less insulating) of the compounds. Recently, we have reported MD coupling in $Ba_3NdRu_2O_9$, suggesting interactions between Ru-*4d* orbitals and *R-4f* orbitals, which is a rare demonstration of MD coupling of a *4d*-orbital based magnetic system.[36] We indicate that such properties are made possible by the particular 6H perovskite structure of the $Ba_3RRu_2O_9$ oxides (Fig.1),[37–40] which consists of isolated $Ru_2O_9$ dimers of two face-sharing $RuO_6$ octahedra interconnected through $RO_6$ octahedra, in this way hindering electronic delocalization in the whole framework.

Considering the significant effects of *R* ions on the complex magnetic ordering and ME/MD couplings previously discovered for *3d* transition-metal based compounds, it is of high interest to investigate the effect of *R*-ion (*4f* orbital) on MD/ME coupling in the $Ba_3RRu_2O_9$ *4d*-orbital based system. In fact, the MD coupling in $Ba_3NdRu_2O_9$ is rather complex and controlled by two different mechanisms, below two magnetic anomalies, at 25 K (ferromagnetic ordering of Nd moments) and 17 K (antiferromagnetic ordering of Nd moments and $Ru_2O_9$ dimers), respectively.[36,41] Motivated by this observation, we have investigated three other members of the $Ba_3RRu_2O_9$ series with *R*=Ho, Tb and Sm. These *R*-ions are selectively chosen from lanthanide series with respect to ionic radii, valence states and magnetic moments (see Table-1), bearing in mind that in those oxides, Ho / Sm are trivalent and Tb is tetravalent, such that the ruthenium dimers have mixed valence $Ru^{4+}/Ru^{5+}$ and single valence $Ru^{4+}$ states, respectively. [37–40] Unlike the aforementioned Nd-based compound, all these three compounds exhibit a single magnetic anomaly around 10-12 K.[37–41]

**Table1: Valence state, ionic radius, spin and orbital moments, and magnetic ordering temperatures Tc for different R-ions of the series Ba₃RRu₂O₉.**

| Lanthanide (R)-ions and valance states | Ionic Radius (pm) | F configurations and effective spin and orbital moments of R-ions | Effective moment of R [$g_J \times (J(J+1))^{0.5}$] ($\mu_B$) | Tc/T$_N$ |
|---|---|---|---|---|
| Ce$^{+4}$ | 101 | $f^0$, S=0 | 0 | No |
| Pr$^{+4}$ | 99 | $f^1$, S=1/2, J=5/2 | 2.54 | 10.5 K |
| Nd$^{+3}$ | 112.3 | $f^3$, S=3/2, J=9/2 | 3.62 | 25 and 17 K |
| Sm$^{+3}$ | 109.8 | $f^5$, S=5/2, J=5/2 | 0.845 | ~ 11 K |
| Eu$^{+3}$ | 108.7 | $f^6$, S=3, J=0 | 0 | ~ 8 K |
| Gd$^{+3}$ | 107.8 | $f^7$, S=7/2, J=7/2 | 7.94 | 14.8 K |
| Tb$^{+4}$ | 106.3 | $f^7$, S=7/2, J=7/2 | 7.94 | 9.5 K |
| Ho$^{+3}$ | 104.1 | $f^{10}$, S=2, J=8 | 10.6 | 10.2 K |
| Er$^{+3}$ | 103 | $f^{11}$, S=3/2, J=15/2 | 9.6 | 6 K |
| Yb$^{+3}$ | 100.8 | $f^{13}$, S=1/2, J=7/2 | 4.54 | 4.5 K |

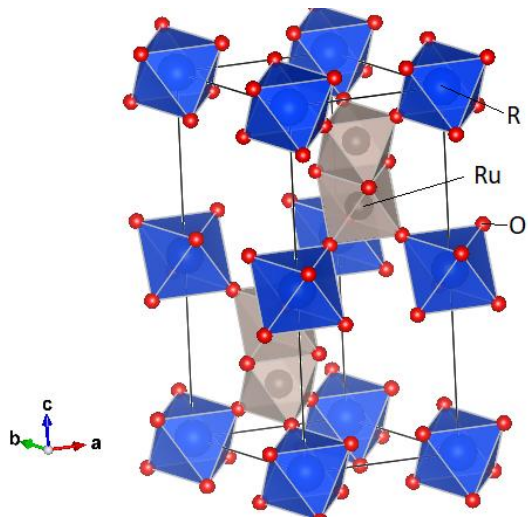

**Figure 1: Crystal structure of Ba₃RRu₂O₉. For clarity, the Ba-atoms are not shown. Two face-sharing distorted RuO₆ octahedra (forming a Ru₂O₉ dimer) and corner sharing RO₆ octahedra are shown.**

## II. Experimental Details

The series of oxides $Ba_3RRu_2O_9$ was synthesized by solid-state-reaction using mixtures of high purity (>99.9%) precursors $BaCO_3$, $RuO_2$ and $R_2O_3$ for $R=$ Ho and Sm and $Tb_4O_7$ for $R=$Tb. These samples were prepared in the form of pellets from intimately mixed powders heated and sintered at temperatures ranging from 1173 to 1573 K with several intermediate grindings, as reported earlier.[37–39,42] The thus-grown samples form a single-phase with the expected P6$_3$/mmc space group in agreement with previous literature.[37–39,42] DC magnetization ($M$) measurements were performed using a Superconducting Quantum Interference Device (SQUID, Quantum Design) as function of temperature ($T$) and magnetic field ($H$). Both temperature and magnetic-field dependent measurements of the complex dielectric measurements with a 1 V ac bias were carried out using a LCR meter (Agilent 4284A) with a home-made sample probe, which is integrated into the Physical Properties Measurement System (PPMS, Quantum Design). Silver paint was used to make parallel-plate capacitors of the pressed disc-like polycrystalline samples (5 mm diameter and 0.7-0.9 mm thickness for different R-members). Positive-up-negative-down (PUND) measurements were performed in a close–loop refrigerator (Janis) using a TF2000 Analyzer equipped with a high voltage booster (Trek 609C) to check for ferroelectric polarization.

## III. Results

### A. Ba$_3$HoRu$_2$O$_9$

The $T$-dependence of the dc magnetic susceptibility of this compound as function of different magnetic fields is shown in figure 2a. For 100 Oe external field, the dc magnetization increases with decreasing temperature and exhibits a clear peak at $T_N \sim$ 10.5 K, indicative for antiferromagnetic (AFM) order which agrees with the previous report by Doi, $et\ al.$[38] The magnetization does not change significantly by increasing the applied magnetic field up to 10 kOe. The application of an external magnetic field of 30 kOe shifts the AFM-type peak to lower temperatures (~8.8 K), as shown in figure 2a. For even higher magnetic fields of 50 kOe, the AFM peak is suppressed and the magnetization becomes almost constant below ~6 K (figure 2a). This decrease of the magnetic-ordering temperature on decreasing temperature is consistent with the proposed AFM behavior. However, such a constant $M(T)$ behavior is not a typical characteristic of a pure AFM system. Figure 2b shows the isothermal magnetization ($M(H)$) at selected temperatures. The isothermal $M(H)$ curve below the magnetic ordering temperature (e.g., at 2 and 7 K) exhibits a linear slope below 30 kOe, like a prototype antiferromagnet. For fields from 30 to 40 kOe, a clear change of slope is observed, unlike a typical AFM system. The inset of figure 2a shows a enlarged plot of $M(H)$ at 2 K, where a weak hysteresis around 30 kOe is observed. The (sudden) increase in magnetization as a function of magnetic field correspond to a $H$-induced magnetic transition from AFM behavior. Such a feature might be referred to as meta-magnetic-type transition, keeping in mind that the sharp feature (step-like increase) around transition is probably smeared out due to polycrystalline nature. Though further spectroscopic investigation is needed to characterize the exact nature of magnetic transition and to understand the change in magnetic structure. The *M(H)* curve at 7 K does not resolute the hysteresis due to its very weak nature. The *M(H)* evolution at *T*= 12 K and 20 K is consistent with the paramagnetic nature of this system. Therefore, we conclude that this system undergoes only one AFM ordering-transition below *T_N*, unlike the Nd-member in this series. The magnetic behavior as documented in Fig. 2b could indicated weak ferromagnetism, such as a canted AFM, or a more complex spin structure due to the presence of competing FM-AFM interactions. Further studies are necessary to elucidate this behavior in full detail. The latter may arise from the ordering of different magnetic ions due to the dominating 4*d*(Ru)-4*f*(*R*) magnetic correlation compared to the 4*d*-4*d* correlation. Note that we did not observe any further ordering at lower temperatures down to 2 K for the Ho-member. The neutron diffraction on $Ba_3NdRu_2O_9$ shows ferromagnetic ordering of Nd at 24 K and canted AFM ordering of Nd-moments below 18 K with simultaneous ordering of the $Ru_2O_9$ dimers, where the Nd-moments are aligned along the c-axis with a small tilting towards the ab-plane and with Ru-moments aligned within the ab-plane.[41] It is not clear whether the ordering at *T_N* reflects the ordering of the Ho-moments and the application of high magnetic fields (*H*>30 kOe) cants the Ho-moments, whereas, $Ru_2O_9$ orders at lower temperature. Another possibility is that both Ru and Ho-moments start to order at *T_N* and the application of magnetic fields further modifies the spin structure and the system stabilizes in a canted magnetic structure.

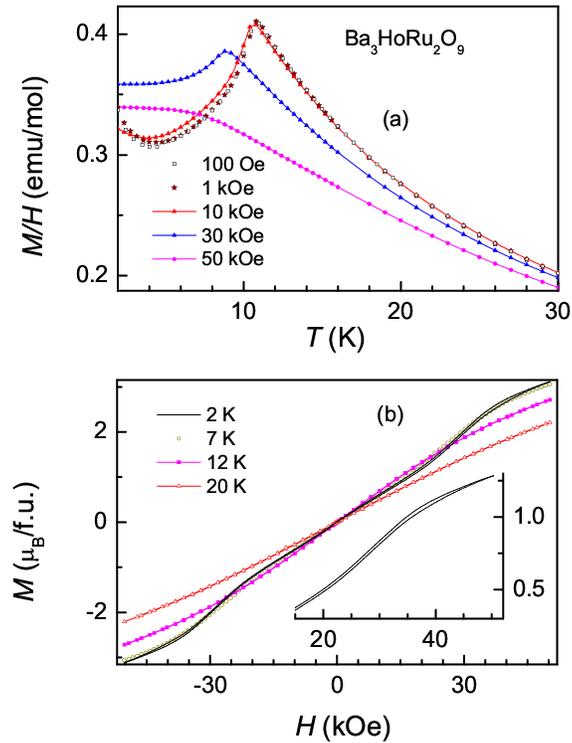

**Figure 2. (a) dc magnetic susceptibility M/H as a function of temperature for a series of magnetic fields ranging from 100 Oe to 50 kOe. and (b) isothermal magnetization *M* at selected temperatures (2 - 20 K) for the compound $Ba_3HoRu_2O_9$. The inset shows an enlarged plot of *M(H)* at 2 K documenting the weak AFM hysteresis.**

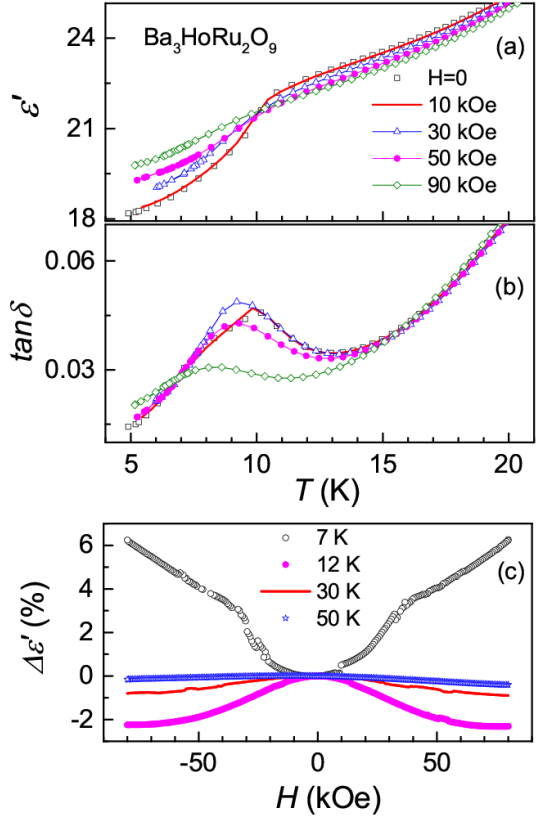

**Figure 3. (a)** Real part of dielectric constant and **(b)** loss tangent as a function of temperature for different magnetic fields for a fixed frequency of 71.4 kHz for the compound Ba$_3$HoRu$_2$O$_9$. **(c)** The excess dielectric constant $\Delta\varepsilon'$ (= [$\varepsilon'(H)$- $\varepsilon'(0)$]/ $\varepsilon'(0)$) as a function of magnetic field for 71.4 kHz at selective temperatures.

Figures 3a and 3b show the real part of dielectric constant ($\varepsilon'$) and the loss tangent $tan\delta$, respectively, as a function of temperature for a fixed frequency of 71.4 kHz in the presence of different magnetic fields. Both $\varepsilon'$ and $tan\delta$ exhibit clear features (kinks and peaks) at the onset of magnetic ordering. No changes of the complex dielectric constant are observed for applied magnetic fields below 10 kOe. Like in the magnetic measurements, the observed dielectric feature at $T_N$ shifts to lower temperature for applied magnetic fields < 30 kOe and broadens for higher magnetic fields (c.f. figure 3a and b for $H$= 50 and 90 kOe). This is in perfect agreement with the results of the magnetic susceptibility, indicating MD coupling. We further confirm this by performing isothermal dielectric measurements (excess dielectric constant, $\Delta\varepsilon'$ = [$\varepsilon'(H)$- $\varepsilon'(0)$]/ $\varepsilon'(0)$), as shown in figure 3c. The magnetic-field dependent changes, $\Delta\varepsilon,'$ remain nearly constant up to ~10 kOe, then quadratically increase with increasing $H$ and finally exhibit a jump above 30 kOe, mimicking the $H$-induced change in $M(H)$. Similar H-induced jump in the dielectric constant is observed in many frustrated multiferroic systems as a result of strong MD coupling exactly near the meta-magnetic transition.[19,21,43,44] No hysteresis is observed in $\Delta\varepsilon'(H)$ as it is very weak at 7 K and even not visible in $M(H)$. Obviously, no such feature is observed above $T_N$. The MD coupling at 7 K ($T<T_N$) is positive, whereas, it is negative at 12 K ($T>T_N$), consistent with $\varepsilon'(T)$ as function of increasing external fields. The negative MD coupling decreases with increasing temperature and becomes nearly zero at further high temperature (see figure 3c). The MD effect above magnetic ordering is due to magnetoelastic coupling, which can be observed in the paramagnetic region as well. The system is highly insulating at low temperature as depicted by the low value of *tanδ*; however, it starts to increases sharply above 15 K. Therefore, the absolute value could be magnified by a small leakage contribution at higher temperature, predicted for the compound $Ba_3NdRu_2O_9$.[36]

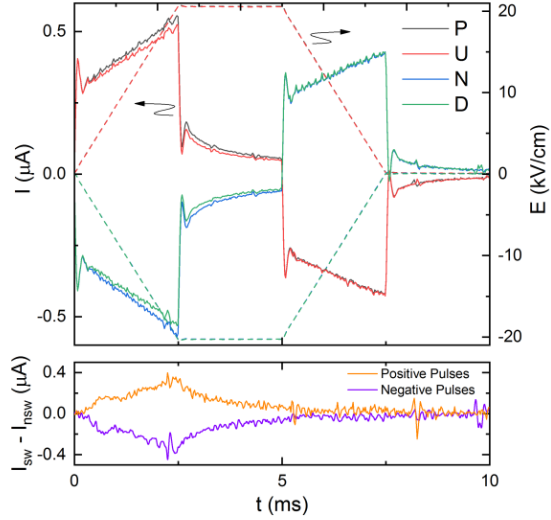

**Figure 4: Results of PUND measurement as discussed in the text. The upper panel shows applied voltage (right axis) and obtained current (left axis) vs. pulse time. P and U corresponds to two consecutive positive pulses with a 1s delay and N and D corresponds to two consecutive negative pulses. Lower panel shows the difference in obtained current between P and U positive pulses and N and D negative pulses.**

To check for ferroelectric polarization we performed non-linear measurements, i.e. PUND measurements, at 5 K. The sample was cooled down from 50 K to 5 K with an applied electrical poling field of 4 kV/cm. At 5 K the poling field was switched off and the sample was kept without any electric fields for 10 min to avoid charging effects. In PUND measurements, two consecutive pulses are used with a time delay between them; application of first pulse (switching pulse) orients both intrinsic dipoles (if it is ferroelectric) and extrinsic contributions (arising e.g. from leakage current), which gives changes in output current; when the applied voltage becomes zero, the extrinsic polarization becomes zero, however dipoles due to intrinsic ferroelectric polarization remain ordered; application of a $2^{nd}$ pulse (non-switching pulse with similar amplitude, direction and frequency as the $1^{st}$ one) again order the extrinsic dipoles and the output current arises only from these extrinsic effects; therefore, any difference in current between these two pulses indicates the intrinsic value of polarization. PUND measurement is often better suited to detect small polarizations of a ferroelectric sample excluding all extrinsic effect.[45–47] The result of the PUND measurement, using electrical excitation fields up to 20 kV/cm, does not reveal a typical signature of a proper ferroelectric behavior. However, there are distinct differences between subsequent switching and non-switching pulses (figure 4). From these results, a relaxed remnant polarization in the order of 30 μC/m² can be estimated, indicating an unconventional ferroelectric polarization in this system. Reversing the electric field reverses the polarization that confirms ferroelectricity. Though, the value of polarization is very small, no such polarization is observed for Nd-member in this series.[36]

To find out the role of the *R*-ion, we have pursued MD measurements of the $Sm^{3+}$ member of this series, whose ionic radius falls between $Nd^{3+}$ an and $Ho^{3+}$.

### B. *Ba₃SmRu₂O₉*

The compound $Ba_3SmRu_2O_9$ exhibits a broad anomaly at high temperature (differently from the Nd and Ho-members), but (similar to them) shows long-range magnetic ordering at 12.5 K ($T_N$).[40] However, the application of a high magnetic field (up to 50 kOe) does not modify the magnetic nature of this system, as documented in figure 5, unlike Nd, and Ho- oxides in this series. The observed low magnetic moment may be related to several factors. One is due to the low magnetization value of $Sm^{3+}$. The magnetization of the Sm-member certainly is low with respect to the low magnetic moment of $Sm^{+3}$ (Hund's rule compared to that of other *R*-member in this series (Table 1). In addition, the $Sm^{3+}$ magnetic moment is extremely sensitive and may be strongly reduced due to crystal field effects (e.g. Ref.[48]), compared to the free-ion value given by Hund's rule. Another possibility is that the Sm-moment does not undergo long-range ordering at $T_N$, the $Ru_2O_9$ dimers undergo long-range ordering as observed in $La^{+3}/Pr^{+4}$ (non-magnetic)-member of this series. Our results demand further spectroscopic investigation in this respect.

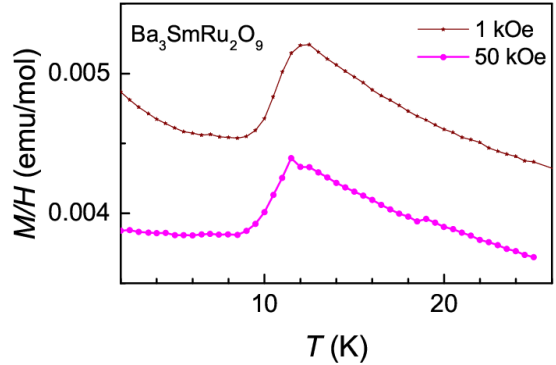

**Figure 5. Dc magnetic susceptibility as a function of temperature for 1 and 50 kOe for the Ba₃SmRu₂O₉.**

We have investigated the dielectric behavior of this Sm-member. We do not observe any clear dielectric feature at the onset of $T_N$ (not shown here), unlike Nd, and Ho-members in this series. No MD coupling is observed (see figure 8).

In an attempt to understand the role of the oxidation states of both *R* and Ru cations and the effect of magnetic moment of *R*-ions on magnetodielectric properties of these oxides, we have investigated $Ba_3TbRu_2O_9$ which exhibits for both cations the tetravalent state, i.e. $Tb^{4+}$ (instead of $R^{3+}$) and $Ru^{4+}$ (instead of $Ru^{4+}/Ru^{5+}$, a feature different from other members.

### C. Ba₃TbRu₂O₉

The compound Ba$_3$TbRu$_2$O$_9$ orders antiferromagnetically at~ 9.5 K,[39] as documented in figure 6a, which was ascribed to antiferromagnetic ordering of the *R*-ions. The ordering temperature decreases with increasing magnetic field, consistent with its AFM nature. Application of a high magnetic field (say, 50 kOe) yields a broad peak, indicating a subtle change on the nature of magnetic ordering. A *H*-induced magnetic transition with a clear magnetic hysteresis at ~ 25 kOe in isothermal *M(H)* below magnetic ordering (see figure 6b) supports such a change in magnetism. The hysteresis around 25 kOe (and no hysteresis around H=0) for Ba$_3$TbRu$_2$O$_9$ is quite clear compared to that of Ba$_3$HoRu$_2$O$_9$. This often arises as a result of 1$^{st}$ order transition, though one cannot rule out the possibility of domain dynamics, specifically in a polycrystalline sample.

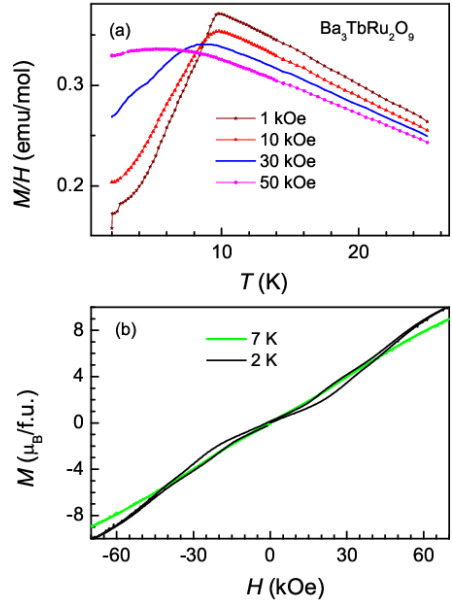

**Figure 6. (a) Dc magnetic susceptibility as a function of temperature for different magnetic field (1-50 kOe) and (b) isothermal magnetization at 2 and 7K for Ba$_3$TbRu$_2$O$_9$.**

Figures 7a and 7b show $\varepsilon'(T)$ and $tan\delta(T)$, respectively, measured at 71.4 kHz in zero field and for various applied magnetic fields. The dielectric anomalies (change in slope in $\varepsilon'$ and peak in $tan\delta$) capture the magnetic ordering around 10 K in zero-magnetic field, which is continuously shifted to low temperatures with increasing magnetic fields (i.e., from 9.5 K ($H = 0$) to 6.5 K ($H = 50$ kOe) and becomes fully suppressed by an applied magnetic field of 90 kOe. At 7 K, the excess dielectric constant $\Delta\varepsilon'(H)$ (figure 7c) exhibits a *H*-induced transition around 25 kOe, as observed in *M(H)*, which sharply increases above 30 kOe with increasing *H*. A small MD coupling is observed above the ordering temperature (figure 7c for *T*= 12 and 30 K), which is consistent with the behavior of $\varepsilon'(T)$ in figure 7a and 7b. Such a small MD coupling may arise from short-range magnetic correlations.

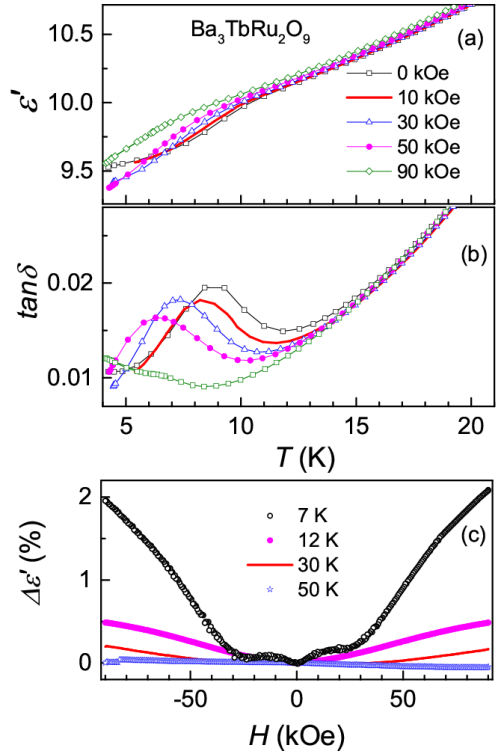

**Figure 7.** (a) Real part of dielectric constant and (b) loss tangent as a function of temperature for different magnetic fields for a fixed frequency of 71.4 kHz for the compound **Ba₃TbRu₂O₉**. (c) The excess dielectric constant *Δε′ (= [ε′(H)- ε′(0)]/ ε′(H))* as a function of magnetic fields measured at 71.4 kHz at selective temperatures for **Ba₃TbRu₂O₉**.

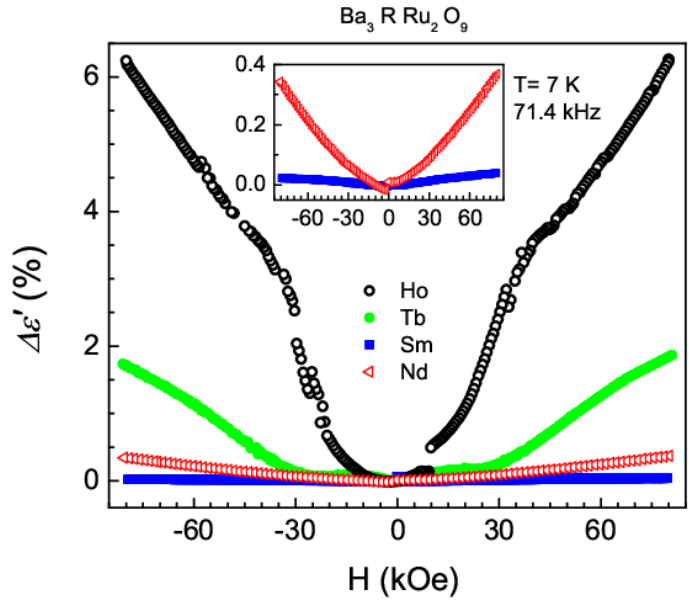

**Figure 8.** The excess dielectric constant *Δε′* as a function of magnetic field for 71.4 kHz at 7 K for *R*=Nd, Sm, Tb and Ho in the series **Ba₃RRu₂O₉**. Inset shows the same plot for the Nd and Sm-members, for better visibility of the MD coupling in these compounds.

## IV.    Discussion and conclusion

This study shows that the *R*=Ho and Tb-oxides exhibit MD coupling similar to what was previously observed for the *R*= Nd-oxide.[36] Importantly, the magnitude of this effect is strongly enhanced, namely by a factor of 3 and 20 for Tb and Ho respectively, as illustrated from the evolution of the excess dielectric constant vs magnetic field at 7 K (Fig. 7). Such a behavior is closely related to the appearance of an AFM (or, canted AFM) transition at 18 K, 12K and 10 K for Nd, Ho and Tb oxides respectively. For the Nd-compound, the magnetic ordering below 18 K originates from an AFM ordering of Ru-Ru FM dimers and canted AFM ordering of Nd.[41] Thus, our study suggests that the MD coupling in those oxides results from the combination of super exchange AFM interactions between $R$O$_6$ octahedra and RuO$_6$ with direct FM interactions between the Ru atoms of the Ru$_2$O$_9$ dimers according to the sequence "$R$-O-Ru-Ru-O-$R$". Within this model, one would expect that the *R*=Sm oxide, which exhibits a $T_N$ of 12K intermediate between the Nd and Ho phases, should also exhibit a MD coupling. This is in contrast with our observation. However, the lanthanide contraction (size effect) is not the only parameter that would account for the MD coupling. In particular the weak magnetic moment of Sm$^{3+}$(0.845 $\mu_B$) compared to other cations Nd$^{3+}$(3.62$\mu_B$), Ho$^{3+}$(10.6$\mu_B$) and Tb$^{4+}$ (7.94$\mu_B$) may explain that for the Sm oxide the signal is too weak to be detected. The increasing value of $\Delta\varepsilon'(H)$ with the magnetic moment following the sequence Sm<Nd<Tb<Ho strongly supports this hypothesis. Such a systematic increase of MD coupling with *R*-ion, characteristic to a systematic increase of the *R*-moments, is a rare observation in this field.  It is difficult to explain the exact mechanism of MD coupling in this 6H-perovskite system without any detailed spectroscopic investigation. The canted antiferromagnetic structure is reported for Nd-members in this series, where the possible role of D-M interaction on positive MD coupling at low temperature is speculated, in addition to possible magneto-striction effect. No such canted AFM behavior in zero magnetic field is reported for other members of this system, though our detailed magnetic results suggest a possible canting above the H-induced magnetic transition for *R*=Tb and Ho. Therefore, D-M interaction from the canted magnetic structure (in presence of high magnetic field at least) probably plays an important role in inducing strong MD coupling. The magnetic exchange energy should be influenced by different extension of rare-earth ions. Interestingly, the magnetic ordering temperature of light rare-earth (Nd)-member is more than two-times higher than that of heavy rare-earth members, in contrast with de-Gennes scaling.[49] In addition, Ba$_3$NdRu$_2$O$_9$ orders ferromagnetically whereas all other rare-earth members order antiferromagnetically. This kind of unusual nature of magnetic ordering for light rare-earth (Nd) compared to that of heavy rare-earths was often attributed to *4f*-hybridization in intermetallic systems (such as, $R_2$PdSi$_3$, see Ref.[50]). However, such *4f*-hybridization is probably difficult to posturize in these highly insulating materials. Magnetic frustration, i.e., exchange frustration due to competing interactions between Ru and *R*-moments, in this complex system could play a significant role on the magnetism and thereby MD coupling. It is to be noted that the *R*-moments are almost aligned parallel to the c-axis whereas the Ru-moments are aligned within the a-b plane at low temperatures (assuming Ba$_3$NdRu$_2$O$_9$ magnetic structure[41]). Therefore, magnetic frustration will be strongly influenced by *4d-4f* magnetic correlations. The strength of magnetic moments and different degrees of hybridization of different $R$-ions both will play a significant role on such strong correlations. It is possible that the stronger magnetic moments of the heavy rare-earths (Ho>Tb>Nd) increase this exchange frustration (due to competing moments of $R$ and Ru). Therefore, heavier rare-earth members undergo long-range ordering at further lower temperature compared to that of Nd-counterpart by minimizing the frustration. This could be consequential to a larger lattice distortion as well to minimize the overall energy which yields a dielectric anomaly at magnetic ordering and strong MD coupling. A weak ferroelectricity is confirmed via PUND measurements for Ho-member in this series, though the absolute value of polarization is very small. However, no signature of ferroelectricity is obtained for light rare-earth Nd-member. This further supports our conclusion. It is possible that this system is multiferroic-I with very small polarization value, where the strong spin-dipole coupling below magnetic ordering arises from higher order coupling term instead of linear magnetoelectric coupling. Such a dominant effect of higher order coupling term below magnetic ordering is demonstrated in the hexagonal multiferroic $RMnO_3$.[10,11] One cannot rule out the small effect of magneto-striction, due to a change in lattice parameter (artifact of geometrical effect of a capacitor). However, this effect cannot be solely responsible for a large MD coupling (say, for Ho-member) since such a huge change in lattice parameter is unlikely (no structural change is observed at the onset of magnetic ordering).

In summary, we have performed a detailed study of both magnetic and magnetodielectric behavior of $Ba_3RRu_2O_9$ oxides for different rare-earth members, selectively chosen from the broad series. A strong enhancement of MD coupling, 3 and 20 times, is demonstrated for heavy rare-earth members, Tb and Ho respectively, compared to the Nd one, which is independent of the valence state of the $R$ and Ru cations. In contrast, no MD coupling is observed for the Sm-phase in spite of the size of $Sm^{3+}$ intermediate between $Nd^{3+}$ and $Ho^{3+}$. This feature suggests that the strength of MD coupling in this system is mainly governed by two parameters which may be antagonist, the size (lanthanide contraction) and the magnetic moment of the rare earth. Detailed neutron investigation on different $R$-members is warranted in order to underpin the exact nature $d$-$f$ correlation and mechanism of MD coupling.

### Acknowledgements

T.B. and A. P. thank the Agence Nationale de la Recherche, France (contract N°ANR- 16-CE08-0023, ANR project LOVE-ME) for financial support of this work. Work at Michigan State University was supported by the U.S. Department of Energy, Office of Science, Office of Basic Energy Sciences, Materials Sciences and Engineering Division under Award # DE-SC0019259.